\documentclass[conference]{IEEEtran}
\IEEEoverridecommandlockouts

\usepackage{url}  
\usepackage{booktabs} 
\usepackage{multirow}
\usepackage{mathtools}

\newcommand\myeq{\stackrel{\mathclap{\normalfont\mbox{def}}}{=}}

\def\BibTeX{{\rm B\kern-.05em{\sc i\kern-.025em b}\kern-.08em
    T\kern-.1667em\lower.7ex\hbox{E}\kern-.125emX}}
\begin{document}

\title{Learning Effective Embeddings From Crowdsourced Labels: An Educational Case Study
\thanks{*Corresponding author: Zitao Liu.}
}

\author{\IEEEauthorblockN{Guowei Xu, Wenbiao Ding}
\IEEEauthorblockA{\textit{TAL AI Lab} \\
Beijing, China \\
\{xuguowei, dingwenbiao\}@100tal.com}
\and
\IEEEauthorblockN{Jiliang Tang}
\IEEEauthorblockA{\textit{Michigan State University} \\
East Lansing, MI USA \\
tangjili@msu.edu}
\and
\IEEEauthorblockN{Songfan Yang, Gale Yan Huang, Zitao Liu\IEEEauthorrefmark{1}}
\IEEEauthorblockA{\textit{TAL AI Lab} \\
Beijing, China \\
\{yangsongfan, galehuang, liuzitao\}@100tal.com} 
}

\maketitle

\begin{abstract}
Learning representation has been proven to be helpful in numerous machine learning tasks. The success of the majority of existing representation learning approaches often requires a large amount of consistent and noise-free labels. However, labels are not accessible in many real-world scenarios and they are usually annotated by the crowds. In practice, the crowdsourced labels are usually inconsistent among crowd workers given their diverse expertise and the number of crowdsourced labels is very limited. Thus, directly adopting crowdsourced labels for existing representation learning algorithms is inappropriate and suboptimal. In this paper, we investigate the above problem and propose a novel framework of \textbf{R}epresentation \textbf{L}earning with crowdsourced \textbf{L}abels, i.e., ``RLL", which learns representation of data with crowdsourced labels by jointly and coherently solving the challenges introduced by limited and inconsistent labels. The proposed representation learning framework is evaluated in two real-world education applications. The experimental results demonstrate the benefits of our approach on learning representation from limited labeled data from the crowds, and show RLL is able to outperform state-of-the-art baselines. Moreover, detailed experiments are conducted on RLL to fully understand its key components and the corresponding performance.
\end{abstract}


\section{Introduction}

It has been suggested that the performance of machine learning applications  strongly relies on the representation of the input data~\cite{bengio2013representation}. A good data representation provides tremendous flexibilities to choose fast and simple models. However, the raw data representation is not typically amenable to learning~\cite{domingos2012few}. Representation learning aims to automatically learn new data representation from raw features by discovering hidden patterns in the data and has attracted increasing attentions in recent years~\cite{bengio2013representation}. Representation learning, especially deep learning~\cite{lecun2015deep} has been immensely advanced the field of machine learning~\cite{goodfellow2016deep} and the related areas such as computer vision~\cite{krizhevsky2012imagenet}, signal processing~\cite{yu2011deep} and natural language processing~\cite{collobert2008unified}. On the other hand, modern successful representation learning approaches are discriminatively trained and often require massive labeled data, which is typically unavailable in many real-world scenarios~\cite{fei2006one}. 

To bridge this gap, human efforts are needed to acquire labeled data manually and crowdsourcing provides a flexible solution. Theoretically, we can obtain a labeled dataset as large as we want via crowdsourcing platforms such as Amazon Mechanical Turk, Figure Eight, etc. However, in practice, the amount of crowdsourced labels for a given task can be {\it limited} due to a variety of reasons. For instance, a limited budget prevents us from affording massive labeled data. Another example is in some domains such as healthcare, privacy concerns or a paucity of data leads to very limited crowdsourced labels. Furthermore, crowd workers are unlikely to be experts and they tend to have different levels of expertise. As a consequence, crowdsourced labels can be very {\it inconsistent}. In other words, two crowd workers can annotate the same object with distinct labels. Given the aforementioned properties of crowdsourced labels, the majority of existing representation learning techniques cannot work appropriately and optimally with crowdsourced labels in practice. 

This problem becomes more critical in building ML models in educational scenarios. The difficulties are two-fold: first, label annotation in educational scenarios usually requires more domain knowledge compared to standard crowdsourcing tasks such as image classification, part-of-speech tagging, etc. It is more ambiguous when labeling a 60-min class (whether the class quality is good or bad) than annotating images. This will lead to very inconsistent labels. Second, labeling each sample in educational scenarios requires much more efforts than standard annotation tasks. For example, it may take a crowd worker less than 1 second to annotate an image while the worker has to watch a 60-min video before determining the class quality.

Recent years have witnessed great efforts on learning with {\it limited} labeled data~\cite{fei2006one,mintz2009distant}. Also inferring true labels from {\it inconsistent} crowdsourced labels has been studied for decades~\cite{dawid1979maximum,ipeirotis2010quality}.  However, research on representation learning with limited and inconsistent crowdsourced labels is rather limited. Thus, in this paper, we study the problem of representation learning with crowdsourced labels. In particular, we target on answering two questions: (1) how to take advantage of crowdsourced labels under the limited and inconsistent settings? and (2) how to build a unified representation learning framework with crowdsourced labels?

In this work, we propose a unified framework, i.e., RLL, to jointly solve problems of learning representation from inconsistent and limited labeled data. In our unified framework, we propose a scheme to generate hundreds of thousands of training instances from only a limited number of labeled data from crowd workers. Furthermore, instead of isolating true label inference from the representation learning process, we use Bayesian inference to estimate the label confidence and integrate the confidence estimation process into the model learning. Our framework's effectiveness is demonstrated in two real-world scenarios with very limited crowdsourced labels. Further experiments are conducted to fully understand the key model components of RLL.

Our major contributions are two-fold: first, we propose a unified framework, i.e., RLL, to jointly solve problems of learning representation from inconsistent and limited labeled data. Second, we conduct experiments on crowdsourced labels in two real-world education applications to fully understand the effectiveness of RLL. 


\section{Related Work}

\subsection{Learning with limited Labeled Data}

Representation learning, especially deep learning has largely advanced the field of machine learning and its applications. Such success typically requires a large amount of labeled data, which is usually unavailable in many domains. Various types of techniques have been developed to enable learning with limited labeled data and next we will review representative techniques. One is few shot learning, which aims to learn new concepts from only a few labeled examples~\cite{fei2006one,jadonsiamese}. When large unlabeled datasets are available, techniques have been developed to make use of weak and distant supervision~\cite{ratner2016data} such as higher-level abstractions~\cite{takamatsu2012reducing}, biased or noisy labels from distant supervision~\cite{liu2017soft} and data augmentation~\cite{dosovitskiy2016discriminative}. Another popular technique for learning from limited labels is transfer learning~\cite{pan2010survey}, which aims to apply knowledge learned in the source domain to a related target domain. 


\subsection{Crowdsourced Labels}

Crowdsourcing offers a flexible way to get labeled data for model learning. Due to the fact that crowd workers have different levels of expertise, the crowdsourced labels are often inconsistent, which can compromise practical applications \cite{sheng2008get}. Therefore,  one key problem is to infer true labels from crowdsourced labels. An EM algorithm is proposed to estimate the error rates when patients answer medical questions with repeated but conflicting responses \cite{dawid1979maximum}. Inspired by Dawid and Skene \cite{dawid1979maximum}, Whitehill et al. considered item difficulty for image classification and a score for each annotator is extracted to assess the quality of the annotator \cite{whitehill2009whose}. Aforementioned approaches can infer the true labels independently, which can be sub-optimal for the targeted tasks. Hence, there are increasing attention on combining true label inference with the targeted machine learning tasks. Raykar et al. proposed an EM algorithm to jointly learn the levels of annotators and the regression models \cite{raykar2010learning}. Likewise, there are efforts to embed label inference process into other types of models. Rodrigues, Pereira, and Ribeiro generalized Gaussian process classification to consider multiple annotators with diverse expertise \cite{rodrigues2014gaussian}. Rodrigues et al. studied supervised topic models for classification and regression from crowds \cite{rodrigues2017learning}. Albarqouni et al. introduced an additional crowdsourcing layer to embed the data aggregation process into convolutional neural network learning \cite{albarqouni2016aggnet}. Very recently, techniques have been studied, which do not need iterative EM algorithms to estimate weights of the annotators. Guan et al. captured information about the annotators by modeling each annotator individually and then learning combination weights via back propagation \cite{guan2017said}. 

The majority of aforementioned algorithms have been designed to address the problems of noise and inconsistencies in crowdsourced labels and they cannot work as expected when labels are limited, especially for these algorithms developed for deep learning. While in this work, we aim to develop algorithms which can jointly solve the challenges from limited and inconsistent labels.  


\section{Methodology}

In this section, we will give details about the proposed framework, which aims to jointly address the challenges from limited and inconsistent crowdsourced labels. Before that, we will introduce notations. 

In this work, we denote the feature vector of \emph{i}th example as $\mathbf{x}_i$ and its corresponding crowdsourced labels are $y_{i,1}, y_{i,2}, \cdots, y_{i,d}$, where $d$ is the number of crowd workers to annotate each sample. Without loss of generality, we assume crowdsourced labels are binary, i.e., $y_{i,j} \in \{0, 1\}$. We use $(\cdot)^+$ and $(\cdot)^-$ to indicate positive and negative examples. We represent the entire data $\mathcal{D}$ as a collection of positive ($\mathcal{D}^+$) and negative examples ($\mathcal{D}^-$). 

The proposed representation learning framework RLL can learn embeddings from limited training data with crowdsourced labels. It is composed of two key components: 

\begin{itemize}
\item \emph{a grouping deep architecture} that learns effective representations from very limited training data.
\item \emph{a Bayesian confidence estimator} that captures the inconsistency among crowdsourced labels and uses Bayesian inference to integrate the label  confidence into the learning process. 
\end{itemize}

\subsection{Grouping Based Deep Architecture}

In many real-world crowdsourcing application scenarios, annotated labels from the crowds are very limited due to many reasons such as budget constraints, privacy concerns, etc. This may easily lead to the overfitting problems for many deep representation models and make them inapplicable. To address this issue, instead of directly training the discriminative representation models from the small amount of annotated labels, we develop a grouping based deep architecture to re-assemble and transform limited labeled examples into many training groups. We would like to include both positive and negative examples into each group. Within each group, we maximize the conditional likelihood of one positive example given another positive example and at the same time, we minimize the conditional likelihood of one positive example given several negative examples. Different from traditional metric learning approaches that focus on learning distance between pairs, our approach aim to generate a more difficult scenario that considers not only the distances between positive examples but distances between negative examples.

 More specifically, for each positive example $\mathbf{x}_i^+$, we select another positive example $\mathbf{x}_j^+$ from $\mathcal{D}^+$, where $\mathbf{x}_i^+ \neq \mathbf{x}_j^+$. Then, we randomly select $k$ negative examples from $\mathcal{D}^-$, i.e., $\mathbf{x}_1^-, \mathbf{x}_2^-, \cdots, \mathbf{x}_k^-$. After that, we create a group $\mathbf{g}_i$ by combining the positive pair and the $k$ negative examples, i.e., $\mathbf{g}_i = <\mathbf{x}_i^+, \mathbf{x}_j^+, \mathbf{x}_1^-, \mathbf{x}_{2}^-, \cdots, \mathbf{x}_{k}^->$. By using the grouping strategy, we can create $O(|\mathcal{D}^+|^2 \cdot |\mathcal{D}^-|^k)$ groups for training theoretically, where $|\mathcal{D}^+|$ and $|\mathcal{D}^-|$ are the number of positive and negative examples in the original labeled data. Let $\mathcal{G}$ be the entire collection of  groups, i.e., $\mathcal{G} = \{\mathbf{g}_1, \mathbf{g}_2, \cdots, \mathbf{g}_n\}$ where $n$ is the total number of groups.

After the grouping procedure, we treat each group $\mathbf{g}_i$ as a training example and feed $\mathbf{g}_i$s into a typical deep neural network (DNN) for learning robust embeddings. The input to the DNN is raw features extracted from each example and the output of the DNN is a low-dimensional semantic feature vector. Inside the DNN, we use the multi-layer fully-connected non-linear projections to learn the compact representations as shown in Figure~\ref{fig:overview}. 

\noindent \textbf{Model Learning} Inspired by the discriminative training approaches in language processing and information retrieval, we propose a supervised training approach to learn our model parameters by maximizing the conditional likelihood of retrieving positive example $\mathbf{x}_j^+$ given positive example $\mathbf{x}_i^+$ from group $\mathbf{g}_i$. 


More formally, let $\mathbf{f}_*$ be the learned representation of $\mathbf{x}_*$ from DNN, where $\mathbf{x}_* \in \mathbf{g}_i$. Similarly, $\mathbf{f}_i^+$ represent the embeddings of the positive example $\mathbf{x}_i^+$. Then, the semantic relevance score between two representations in the embedding space within a group is measured as $r(\mathbf{x}_i^+, \mathbf{x}_*) \myeq  \mbox{cosine}(\mathbf{f}_i^+, \mathbf{f}_*)$.

In our representation learning framework, we compute the posterior probability of $\mathbf{x}_j^+$ in group $\mathbf{g}_i$ given $\mathbf{x}_i^+$ from the cosine relevance score between them through a softmax function

\begin{equation}
\label{eq:softmax}
p(\mathbf{x}_j^+|\mathbf{x}_i^+) = \frac{\exp \big(\eta \cdot r(\mathbf{x}_i^+, \mathbf{x}_j^+)\big)}{\sum_{\mathbf{x}_* \in \mathbf{g}_i, \mathbf{x}_* \neq \mathbf{x}_i^+} \exp \big(\eta \cdot r(\mathbf{x}_i^+, \mathbf{x}_*)\big) } \nonumber
\end{equation}

\noindent where $\eta$ is a smoothing hyper parameter in the softmax function, which is set empirically on a held-out dataset in our experiment. 

Hence, given a collection of groups $\mathcal{G}$, we optimized the DNN model parameters by maximizing the sum of log conditional likelihood of finding a positive example $\mathbf{x}_j^+$ given the paired positive example $\mathbf{x}_i^+$ from group $\mathbf{g}_i$, i.e., $\mathcal{L}(\Omega) = - \sum_{i=1}^{n} \log p(\mathbf{x}_j^+|\mathbf{x}_i^+)$, where $\Omega$ is the parameter set of the DNN. Since $\mathcal{L}(\Omega)$ is differentiable with respect to $\Omega$, we use gradient based optimization approach to train the DNN.

\subsection{Bayesian Confidence Estimator}

When obtaining labels from the crowds, each example is usually labeled by $d$ workers and by nature the crowdsourced labels are not consistent. Moreover, it is unrealistic to hire a large amount of crowd workers to label the same example multiple times to avoid the inconsistency via majority vote. In most cases, even though two examples are both identified as positive, the confidence of their ``positiveness'' might be different. For example, assuming $\mathbf{x}_1^+$ and $\mathbf{x}_2^+$ are two positive examples whose corresponding 5-person crowdsourced labels are $(1, 1, 1, 1, 1)$ and $(1, 1, 1, 0, 0)$. Apparently, our assurance of their labels should be different and we should consider such crowdsourced labeling inconsistency into our model training process. In this work, we model such inconsistency as the confidence about the crowdsourced labels. Let $\delta_i$ be the confidence of each example $\mathbf{x}_i$.

One obvious approach is to treat the confidence ($\delta_i$) as a random variable that follows the Bernoulli distribution and obtain $\delta_i$ by using maximum likelihood estimation (MLE) as follows:

\vspace{-0.4cm}
\begin{equation}
\label{eq:mle}
\delta_i^{\mbox{MLE}} = \sum_{j=1}^d y_{i,j} / d
\end{equation}

However, in many real-world setting, we are not able to afford too many crowd workers to label the same example simultaneously, i.e, \emph{d} is relatively small. This leads to the inferior performance in the MLE approach (eq.(\ref{eq:mle})). To address this problem, we assign a Beta prior to $\delta_i$, i.e., $\delta_i \sim \mbox{Beta}(\alpha, \beta)$. Therefore, the posterior estimation of the crowdsourced label confidence is 

\vspace{-0.4cm}
\begin{equation}
\label{eq:posterior}
\delta_i^{\mbox{Bayesian}} = \frac{\alpha + \sum_{j=1}^d y_{i,j}} {\alpha + \beta + d}
\end{equation}

After that, we integrate the crowdsourced label confidence into our representation learning. The confidence weighted conditional probability is defined as follows:

\vspace{-0.4cm}
\begin{equation}
\label{eq:softmax_2}
\hat{p}(\mathbf{x}_j^+|\mathbf{x}_i^+) = \frac{\exp \big(\eta \cdot \delta_j \cdot r(\mathbf{x}_i^+, \mathbf{x}_j^+)\big)}{\sum_{\mathbf{x}_* \in \mathbf{g}_i, \mathbf{x}_* \neq \mathbf{x}_i^+} \exp \big(\eta \cdot \delta_* \cdot r(\mathbf{x}_i^+, \mathbf{x}_*)\big) }
\end{equation}

\noindent where $\delta_j$ and $\delta_*$ are confidence scores of $\mathbf{x}_j^+$ and $\mathbf{x}_*$. Accordingly, we adjust the objective function by using the confidence weighted conditional probability (eq.(\ref{eq:softmax_2})).

\subsection{Model Summary}

\begin{figure*}[!tpbh]
\centering
\includegraphics[width=0.75\textwidth] {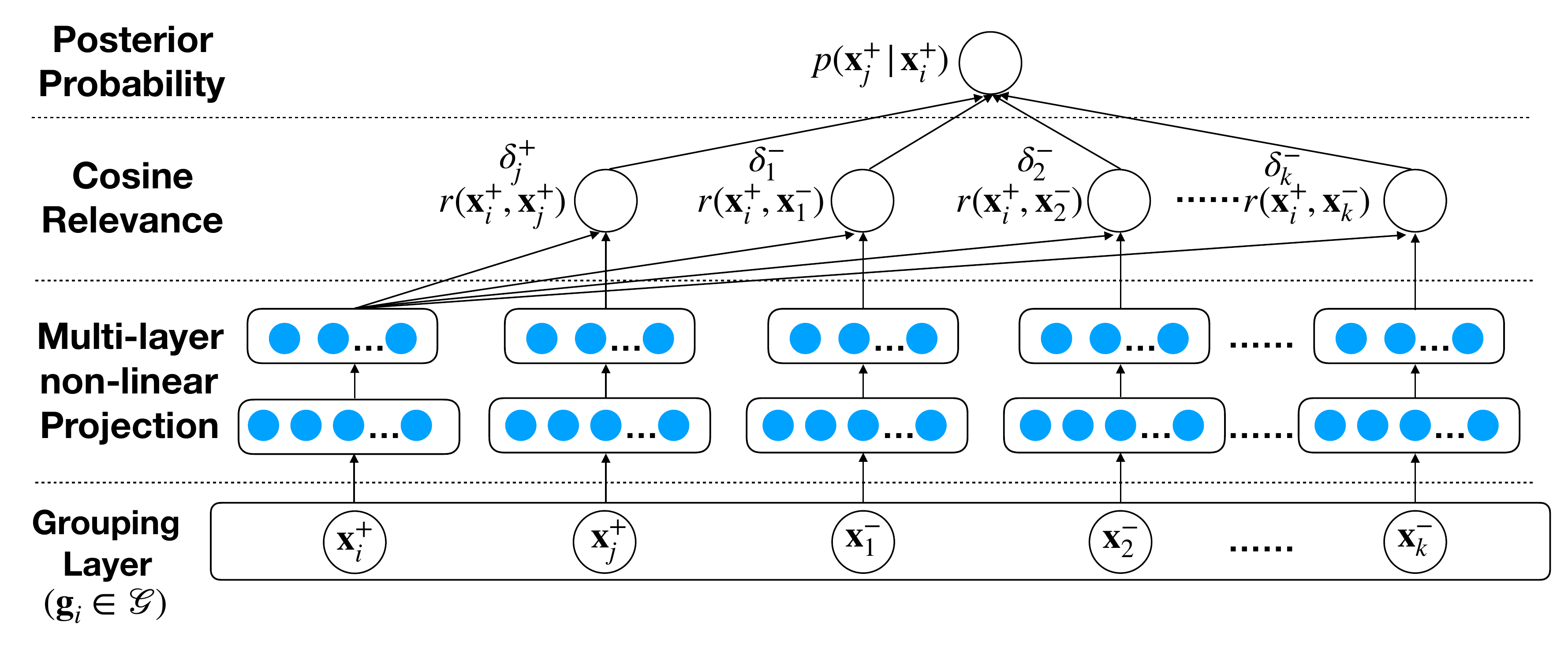}
\caption{The overview of the RLL framework.}
\label{fig:overview}
\end{figure*}

In our RLL framework, given the limited data with crowdsourced labels, we first generate a fair large amount of groups of training examples by including both positive and negative examples. Then, we estimate the label confidence for each crowdsourced data by a Bayesian estimator. After that, we feed all groups into a DNN which maximizes the confidence-weighted conditional likelihood of retrieving the positive examples. The entire RLL framework is illustrated in Figure \ref{fig:overview}. 

\section{Experiments}

In this section, we evaluate our approach on two real-world crowdsourced datasets. We first introduce the experimental setting, next validate the effectiveness of the proposed framework by comparing with representative baselines and finally study the important parameters of the proposed framework. To encourage the reproducible results, we make our code publicly available at: \url{https://github.com/tal-ai/Representation-Learning-with-crowdsourced-Labels}.

\subsection{Experimental Setting}

To assess the proposed framework, we conduct several experiments on two real-world datasets. 

\begin{itemize}
\item {\bf Oral Math Questions}(``oral'') We collect 880 audio files about oral math questions from students in grade 2. In each wav, a student talks about his or her entire thinking process of solving a math question. Our task is to predict whether the student's entire speech is fluent. 
\item {\bf Online 1v1 Class Qualities} (``class'') We collect 472 videos from public online 1v1 classes where a teacher is teaching classes to a student on a live broadcast platform. The task is to predict the quality of the entire class based on the interactions between the teacher and the student, whether the students take notes, etc. The average length of our class video files is 65 minutes.
\end{itemize}

In this work, each example in both \emph{oral} and \emph{class} are annotated by five annotators. The annotation tasks for both datasets are binary classification tasks in which the workers only need to assign 1 or 0 to instances. 1 represents fluent and good quality in \emph{oral} and \emph{class} respectively. Please note that data annotation for multimedia data (audio and video data) is very time-consuming. For example, to label an example in \emph{class}, the worker has to watch the entire 65-minute video before giving the label. In addition, experts are also asked to annotate these two datasets and expert labels are considered as the ground truth for evaluation purpose. For expert labels, in \emph{oral} dataset, the positive (students' speeches are fluent) over negative (students' speeches are influent) samples ratio is 1.8. In \emph{class} dataset, the positive (the class is in good quality) over negative (the class is not in good quality) sample ratio is 2.1.

Following the tradition to assess representation learning algorithms, we evaluate the classification performance via accuracy, and F1 score. We choose logistic regression as the basic classifier. For each task, we conduct a 5-fold cross validation on the datasets and report the average performance. We use label class prior to set the hyper parameters $\alpha$ and $\beta$ in eq.(\ref{eq:posterior}) for both \emph{oral} and \emph{class}.

\subsection{Performance Comparison}

To assess the effectiveness of the proposed framework, we carefully choose three groups of state-of-the-art as our baselines. 

\subsubsection{Group 1: True Label Inference from Crowdsourcing}

The first group contains methods inferring true labels from crowdsourced labels. For examples in both \emph{oral} and \emph{class} datasets, we extracted a wide range of linguistic features from the raw texts after having automatic speech recognition on the videos. They are listed as follows:

\begin{itemize}
\item Logistic regression with every pair (instance, label) provided by each crowd worker as a separate example. Note that this amounts to using a soft probabilistic estimate of the actual ground truth to learn the classifier, i.e., \emph{SoftProb} \cite{raykar2010learning}. 
\item Logistic regression with EM labels, i.e., \emph{EM} \cite{dempster1977maximum}. The labels are treated as hidden variables and inferred by expectation-maximization. The EM algorithm iteratively estimated worker's accuracy and exploited the estimated accuracy to compute the aggregated result.
\item Logistic regression with GLAD labels, i.e., \emph{GLAD}. GLAD infers the true labels by jointly inferring the true label, worker's expertise and the difficulty of each data instance \cite{whitehill2009whose}.
\end{itemize}

\subsubsection{Group 2: Representation Learning with Limited Labels}

The second group includes representation learning methods designed for limited labels. We use the majority vote from the crowdsourced labels to infer the true labels. They are listed as follows:

\begin{itemize}
\item Siamese networks, i.e., \emph{SiameseNet} \cite{koch2015siamese}. We train a siamese network that takes a pair of examples and trains the embeddings so that the distance between them is minimized if they're from the same class and is greater than some margin value if they represent different classes.
\item Triplet networks, i.e., \emph{TripleNet} \cite{schroff2015facenet}. We train a triplet network that takes an anchor, a positive (of same class as an anchor) and negative (of different class than an anchor) examples. The objective is to learn embeddings such that the anchor is closer to the positive example than it is to the negative example by some margin value.
\item Relation network for few-Shot learning, i.e., \emph{RelationNet} \cite{yang2018learning}. The RelationNet learns a deep distance metric to compare a small number of images within episodes. 
\end{itemize}

\subsubsection{Group 3: Two-stage Models by Combining Group 1 and Group 2}

The third group are methods combining baselines from the first (i.e., inferring true labels) and second groups (i.e., learning embedding with limited labels). They solve the problems of the limited and inconsistent labels in two stages. Due to the page limit, we only combine the best approaches from first and second groups. 

\subsubsection{Group 4: Our Methods}

We also create some variants of our RLL framework, i.e., RLL, RLL-MLE and RLL-Bayesian, as follows:

\begin{itemize}
\item Learning representation by using RLL without Bayesian confidence score and true labels inferred from majority vote, i.e., RLL. 
\item Learning representation by using RLL with confidence score estimated by MLE, i.e., RLL-MLE.
\item Learning representation by using RLL with confidence score estimated by Bayesian approach, i.e., RLL-Bayesian.
\end{itemize}

\subsection{Experimental Results}

The accuracy and F1 score performance on the {\it oral} and {\it class} dataset are demonstrated in Table \ref{tab:predict}. From these results, we make the following observations: 

\begin{itemize}
\item Methods in {\bf group 3}, which solve the problem of limited and inconsistent labels by combining baselines from {\bf group 1} and {\bf group 2}, tend to obtain better performance than the corresponding individual baselines that address either only inconsistent labels in {\bf group 1} or only limited labels in {\bf group 2}.  These results suggest that solving the two problems about crowdsourced labels together is necessary and can benefit the applications which make use of crowdsourced labels. 
\item The proposed frameworks perform better than methods in {\bf group 3}. Methods in {\bf group 3} are two-stage algorithms; while our representation framework frameworks jointly solve the problems from limited and inconsistent labels in a unified and coherent manner.  
\item  RLL-Bayesian always outperforms RLL-MLE and RLL. Crowdsourcing labels usually are inconsistent due to the different backgrounds and education levels of crowd workers. Meanwhile, it is unrealistic and impractical to solve the inconsistency issue by hiring an ``infinite'' group of crowd workers. As we can see from the results, both RLL-Bayesian and RLL-MLE show better performance compared with RLL, which indicates the necessity to consider the confidence of crowdsourcing labels when learning from crowdsourced labels. Comparing RLL-MLE and RLL-Bayesian, RLL-Bayesian achieves better performance. Because of the limited number of crowdsourced labels for each example, the label confidence estimation cannot purely rely on MLE and we should utilize the prior knowledge to guide the confidence estimation. 
\item Performance of methods in {\bf group 3} vary a lot and sometimes are much worse than methods in {\bf group 1}. Modern representation learning approaches such as SiameseNet rely heavily on the complex neural network and massive training samples. When the training samples become limited, their performance cannot be guaranteed and may easily run into the overfitting problem and could be inferior compared to classic methods in {\bf group 1}.
\end{itemize}

\vspace{-0.3cm}
\begin{table}[ht]
\centering
\scriptsize
 \caption{Prediction results oral and class datasets.}
\begin{tabular}{r*{7}{c}}
  \toprule
  && \multicolumn{2}{c}{Oral Data} & \multicolumn{2}{c}{Class Data} \\
  \cmidrule(lr){3-4} \cmidrule(lr){5-6}
  Method & Group & Accuracy & F1 & Accuracy & F1 \\
  \midrule
  SoftProb           	& group 1 & 0.815 & 0.869 & 0.758 & 0.810 \\
  EM                  	& group 1 & 0.843 & 0.887 & 0.606 & 0.698 \\
  GLAD                	& group 1 & 0.831 & 0.881 & 0.697 & 0.773 \\ \midrule
  SiameseNet        	& group 2 & 0.802 & 0.859 & 0.719 & 0.836 \\
  TripletNet          	& group 2 & 0.847 & 0.889 & 0.750 & 0.857 \\
  RelationNet         	& group 2 & 0.843 & 0.890 & 0.730 & 0.842 \\ \midrule
  SiameseNet+EM     & group 3 & 0.798 & 0.856 & 0.727 & 0.842 \\
  SiameseNet+GLAD & group 3 & 0.815 & 0.871 & 0.727 & 0.842 \\
  TripletNet+EM         & group 3& 0.843 & 0.887 & 0.727 & 0.842 \\
  TripletNet+GLAD     & group 3& 0.843 & 0.890 & 0.667 & 0.792 \\
  RelationNet+EM      & group 3 & 0.860 & 0.899 & 0.727 & 0.842 \\
  RelationNet+GLAD  & group 3 & 0.854 & 0.889 & 0.730 & 0.842 \\ \midrule
  RLL     			& group 4 & 0.871 & 0.901 & 0.818 & 0.880 \\
  RLL+MLE   		& group 4 & 0.871 & 0.903 & 0.848 & 0.902 \\
  RLL+Bayesian 	& group 4 & \bf{0.888} & \bf{0.915} & \bf{0.879} & \bf{0.920} \\ 
  \bottomrule
\end{tabular}
 \label{tab:predict}
\end{table}

To sum up, the proposed frameworks significantly benefit the classification performance. Next, we design experiments to further understand the working of the proposed frameworks. Since RLL-Bayesian always obtain the the best performance among the three variants, the following investigations are based on RLL-Bayesian.

\subsection{The Impact of Negative Examples}

One important parameter of the proposed framework is the number of negative examples $k$ we include in each group $g_i$. To understand the impact of negative examples, we examine how the performance changes when we vary $k$ as $\{2,3,4,5\}$. The performance variances of the proposed framework w.r.t. $k$ are demonstrated in Table \ref{tab:varied_k_metric} for the \emph{oral} and \emph{class} datasets respectively.  

\vspace{-0.3cm}
\begin{table}[ht]
\centering
 \caption{RLL-Bayesian results with different $k$s.}
\begin{tabular}{r*{5}{c}}
  \toprule
  & \multicolumn{2}{c}{Oral Data} & \multicolumn{2}{c}{Class Data} \\
  \cmidrule(lr){2-3} \cmidrule(lr){4-5}
  $k$ & Accuracy & F1 & Accuracy & F1 \\
  \midrule
  2                 & 0.809 & 0.852 & 0.699 & 0.813 \\
  3                  & \bf{0.888} & \bf{0.915} & \bf{0.879} & \bf{0.920} \\
  4                & 0.831 & 0.875 & 0.757 & 0.855 \\ 
  5                & 0.803 & 0.851 & 0.750 & 0.846 \\ 
  \bottomrule
\end{tabular}
 \label{tab:varied_k_metric}
\end{table}

From Table \ref{tab:varied_k_metric}, we can observe that in general, with the increase of $k$, the performance tends to first increase and then decrease. When $k$ is from $2$ to $3$, the performance increases remarkably. We can have more groups (or training samples) with $k=3$ compared to $k=2$. However, when we increase $k=3$ to $k=4$ and $k=5$, the performance decreases dramatically. Though by increasing $k$, we can get more training samples, we also may introduce noise.

\subsection{Impact of the Number of Crowd Workers}


Another important parameter of the proposed framework is the number of crowd workers $d$ to annotate each data sample. Similar to the analysis on $k$, we check the performance changes by choosing $d$ from $\{1,3,5\}$. The results are demonstrated in Table \ref{tab:varied_d_metric} for \emph{oral} and \emph{class} datasets. As shown in Table \ref{tab:varied_d_metric}, with the increase of $d$, the performance consistently increases. With more crowd workers to label each data instance, our model is more likely to estimate the trustworthiness of crowdsourced labels. The observation can help us determine the reasonable number of crowd workers for crowdsourcing in practice.

\vspace{-0.3cm}
\begin{table}[ht]
\centering
 \caption{RLL-Bayesian results with different $d$s.}
\begin{tabular}{r*{5}{c}}
  \toprule
  & \multicolumn{2}{c}{Oral Data} & \multicolumn{2}{c}{Class Data} \\
  \cmidrule(lr){2-3} \cmidrule(lr){4-5}
  $d$ & Accuracy & F1 & Accuracy & F1 \\
  \midrule
  1                 & 0.826 & 0.873 & 0.727 & 0.842 \\
  3                  & 0.876 & 0.922 & 0.758 & 0.840 \\
  5                & \bf{0.888} & \bf{0.915} & \bf{0.879} & \bf{0.920} \\ 
  \bottomrule
  \end{tabular}
 \label{tab:varied_d_metric}
\end{table}

\section{Conclusion}

In this work, we study the problem of representation learning with crowdsourced labels. We design a novel representation learning framework RLL for crowdsourced labels under the limited and inconsistent settings. Experimental results on two real-world applications demonstrate (1) the proposed framework outperforms the representative baselines; and (2) it is necessary to address the limited and inconsistent label problems simultaneously. Our current model does not make use of any information about individual crowd worker and we want to extend the proposed framework to incorporate such information in the future.

\section*{Acknowledgements}
Jiliang Tang is supported by the National Science Foundation (NSF) under grant numbers IIS-1714741, IIS-1715940 and CNS-1815636, and a grant from Criteo Faculty Research Award.
\bibliographystyle{IEEEtran}
\bibliography{crowdsourcing}

\end{document}